\documentclass[twoside]{article}
\usepackage{fleqn,espcrc2}

\usepackage{graphicx}
\usepackage[figuresright]{rotating}

\newcommand{\AmS}{{\protect\the\textfont2
  A\kern-.1667em\lower.5ex\hbox{M}\kern-.125emS}}

\title {Design and expected performance of the ANTARES neutrino telescope.}

\author{F.Montanet \address{CPPM, 163 av. de Luminy - case 907 , 13288
    Marseille Cedex 9, France} \thanks{Centre de Physique des
    Particules de Marseille, IN2P3-CNRS / Univ. de la
    M\'editerran\'ee}
    (on behalf of the ANTARES Collaboration)
}

\begin{document}

\begin{abstract}
The ANTARES Collaboration is aiming at the construction and the operation of
a large undersea neutrino telescope for neutrino astronomy, neutrino oscillation
and indirect dark matter searches. An intensive R\&D program, which started 3 years ago,
has shown the feasibility of such a detector in the deep waters of the Mediterranean
sea. We have now started the design and the construction of a 0.1 km\( ^{2} \) detector
for which the expected performance will be briefly described here. \vspace{1pc}
\end{abstract}

\maketitle

\section{Aims and principle.}
The ANTARES (Astronomy with a Neutrino Telescope and Abyss environmental RESearch)
is an international collaboration which aims at the construction and the operation
of a large undersea telescope for the detection and the study of high energy
cosmic neutrinos\cite{Proposal99}. The collaboration is rapidly growing and
is composed of particle physicists and of astronomers, as well experts in
sea science and technology.

The physics and astrophysics aims are described in more detail in the contribution
from L. Moscoso to this conference\cite{Luciano-taup99}. The basic idea of
the detection of high energy cosmic neutrino by large undersea or under-ice
Cerenkov detectors is almost 40 years old: an array of optical modules (OMs)
is used to detect the Cerenkov signal emitted by muons in water. Muons are
induced by charged current interaction of neutrinos at some distance of the
detector. The target size is thus of the order of the muon range i.e. much
larger than the detector itself. The overwhelming background of down
going muons produced in the high atmosphere is reduced by shielding the apparatus
under some thousands of meters of water and rejecting the remaining atmospheric
muons by looking only at upward going muons. The muon trajectory is accurately
reconstructed from the Cerenkov photons arrival time information on each photomultiplier
(PMT) contained in the optical modules. At high energy (above a few TeV) the
neutrino direction is well preserved at the interaction and the resulting pointing
accuracy is better than a fraction of a degree allowing accurate source identification.
Given that the expected fluxes are very low, the ultimate goal is the realisation
of a km$^3$ detector which should record hundreds or thousands of cosmic
neutrino events with energies above a few TeV but a 10 times smaller detector 
would be able to reveal first high energy cosmic neutrinos and maybe identify 
some point-like sources. At lower energies, contained
or semi contained events can be used and the neutrino energy can be inferred
from the muon range measurement, giving access to neutrino oscillation physics
using atmospheric neutrinos or to indirect neutralino searches in the core of
the Earth, the Sun or the Galaxy.

\section{The R\&D program.}
Since 1996, the ANTARES collaboration performed an active R\&D program to
show the feasibility of a large undersea detector. Indeed, past experience
had shown that the realisation of such a detector is not trivial and that specific
studies in sea technology were needed. The required studies were carried
out by ANTARES and concerned: the mechanical structure of the elementary
detector lines, the deployment and recovery techniques, the connection of the
detector to the shore with an electro-optical cable for energy supply and data
transfer, PMT front-end electronics, data readout and remote control, the monitoring
of the OM positions on the whole structure. Furthermore, many questions concerning
the deep sea environmental parameters, water optical properties and long term effects 
needed to be assessed. For that reason, ANTARES performed many  
{\em in situ} measurements to answer these questions.

{\bf Environmental studies:}\\
Many measurements and long term survey of environmental
parameters such as current velocity and variation, optical background, light
attenuation and scattering in water, and fouling on OMs were performed. 
These measurements have been obtained {\em in situ} by instrumented autonomous 
mooring lines deployed on a Mediterranean site located off-shore from Toulon (France) 
at a depth of 2400~m (hereafter called the ANTARES site).

The optical background is studied by recording the counting rate of OMs as function
of the time. For a 8'' diameter PMT, it shows a continuous level of less than
50~kHz due to Cerenkov emission from $^{40}$K $\beta$-decays
and spikes with typical duration of one second coming from bioluminescence activity.

Several measurements of the sea water optical properties have been performed
by looking at the arrival time distribution on a small PMT placed 24 or 44~m
away from a pulsed blue LED emitting 466~nm photons. A comparison of the
relative proportion of direct and delayed photons leads to an absorption length
of 55-65~m accounting for seasonal variations and to a scattering length
greater than 200~m for large angle (Rayleigh-like) scattering.

The effect of sedimentation and bio-fouling on the transparency of the optical
surface was monitored over long periods using a setup consisting of 
continuous light source and PIN diodes at different positions on a optical 
module sphere. The optical attenuation was measured to be less than
1.5\% after 8 months.

The ANTARES site is now well studied and the deep water and environmental parameters 
are found to be acceptable for a neutrino telescope.

{\bf Mastering the detector deployment:}\\
It was soon understood that owing to the critical deployment of any large
mechanical structure and to the necessity to reach long term reliability,
the detector structure should be as simple as possible: mere flexible string-like 
mooring lines anchored on the sea bed and held up by buoyancy, supporting optical
modules.
The counterpart of this simplicity is a rather sparse horizontal detector density 
and the necessity to accurately monitor the position of every single element along 
the strings.

In order to learn about the complex deployment procedure
as well as the mechanical behaviour of the detector during the deployment phase
and when it rests at the bottom of the sea, a demonstrator line consisting of a 350~m 
high detector string was designed and built. 
This line is made of two vertical cables, separated
by 2~m, supporting 16 frames each holding a pair of optical modules. The frames are
placed every 15~m, starting 100~m above the sea bed.
It is fully equipped
as far as cabling and electronics containers. The line also contains all the
sensors needed for the precise positioning of the detector elements and for
the recording of the environmental parameters.
Successful deployment tests of this line at a depth of 2400~m have been performed
in summer 1998 using a dynamical positioning ship, showing that the
deployment and recovery procedures were well mastered. It was also
proved that the bottom of a string could be set at its aimed position
on the sea bed with an accuracy of the order of a meter.
The string was then equipped with 8 large dimension PMTs (8 and 10'') as well as 
the electronics needed to transmit the signal to the shore via a 37~km
long electro-optical cable. It has been successfully connected to the shore
station and deployed November 26th 1999 at a depth of 1100~m for
long-term running. Raw background events and atmospheric muon data are 
currently being recorded and analysis is in progress.

In December 1998, we also performed successful tests of undersea electrical
connections of a detector anchor at 2400~m depth using the IFREMER submarine
vehicle ({\it Le Nautile }). 

All this insures the feasibility of the
installation of an array of instrumented strings at the bottom of the sea.

\section{A 0.1 km$^2$ detector.}
Since spring 1999, the ANTARES Collaboration is starting the second phase of
the project which is the design of a 0.1 km$^2$ undersea neutrino 
detector\cite{Proposal99}.

This detector will be equipped with a total of about 1000 OMs placed on 13 mooring
lines of 400~m high and spaced by 60 to 80~m. Each line will be connected
to a junction box using a submarine vehicle, the junction box being connected
to the shore station through a 50~km electro-optical cable. This 13 string
detector is aimed to be deployed on the ANTARES site by 2003. This second phase
of the ANTARES project is already approved by the French and Spanish scientific
councils, it will be decided soon in the UK and the Netherlands.

This 0.1 km$^2$ detector is foreseen to be devoted to three main topics.
The first one is the neutrino astronomy i.e. the study of cosmic neutrinos with
an energy above 1 TeV which may come from diffuse signal, point sources such
as individual AGN. The second subject is the search for
neutrinos coming from dark matter particle annihilations in the centre of the
Earth, in the Sun or in the Galactic centre. The third topic is the study of
atmospheric neutrino oscillations in the 5-100 GeV range.

Extensive simulation studies have been carried out to understand the performances
of such a detector for these different topics 
(see for exemple ICRC proc. references in \cite{Luciano-taup99}). 
%(for details, see references in \cite{Luciano-taup99}). 
For high energy neutrino astronomy,
the angular resolution is a crucial point. Simulations show that
because of the good optical quality of the sea water (low scattering)
and taking into account the good timing capability of the detector, a
point source will be reconstructed with half of the events contained 
within $0.2^\circ$ of the source direction. This result will permit 
the division of the sky map into 200~000~pixels. 
From the amount of light measured by the optical modules, the energy
of the muon is estimated within a factor~3 for muons in the 1-10 TeV
range and within a factor~2 above 10 TeV.  
These performances will allow the detection of a signal of cosmic neutrinos coming
from cosmic sources such as AGNs above the atmospheric neutrino
background: this analysis would be performed by imposing a threshold
on the reconstructed neutrino energy to reduce the atmospheric
neutrino contribution to the diffuse flux or enriching a signal coming
from point sources by overlaying the pixels of the 43 known AGNs
detected by EGRET.

Studying upward going atmospheric neutrino events with an interaction point
inside the detector and using the low energy (5-100 GeV) muon range as
an estimator of the parent neutrino energy, on can explore the
physics of atmospheric neutrino oscillations with mass difference in
the range $\Delta m^2$ between $10^{-3}$ and  $10^{-4}$ eV$^2$. Our
analysisis based on the shape of $E/L$ distribution and is basically
independent of the poorly known absolute $\nu_{atm}$ flux. 
In case of a positive oscillation
signal in $\Delta m^2$-$\sin^22\theta$ region allowed by the
Super-Kamiokande experiment, the fit would lead to a precise
detemination of the oscillation parameters.
On the contrary, in absence of oscillation, the excluded region would
well cover the region allowed by Super-Kamiokande.
 
\section{Conclusions}
The R\&D program performed by the ANTARES Collaboration has
demonstrated that the water and environmental properties of the chosen ANTARES site
are well suited for the installation of the first stage of a large size neutrino
telescope. It was also demonstrated that the necessary marine
technologies concerning aspects such as detector deployment, undersea
connections, positionning, long term reliability, etc, are well under
control.
ANTARES is starting the next step towards the km-scale neutrino telescope by
the design, the installation and the running of a 0.1~km$^2$ detector
off the Mediterranean coast of France by 2003. This detector
will play a pioneering role in neutrino astronomy.

\end{document}